\documentclass[aps,prb,twocolumn,superscriptaddress,showpacs,amssymb,amsmath]{revtex4}
\usepackage{graphicx}
\usepackage{amssymb}
\begin{document}
\title{Relaxation mechanism for electron spin in the impurity band of n-doped 
semiconductors}
\author{Pablo\ I.\ Tamborenea}
\affiliation{Departamento de F\'{\i}sica, FCEN, Universidad de Buenos Aires, 
Ciudad Universitaria, Pab.\ I, C1428EHA Ciudad de Buenos Aires, Argentina}
\affiliation{Institut de Physique et Chimie des Mat\'eriaux de Strasbourg, 
UMR 7504 (CNRS-ULP), 23 rue du Loess, BP 43, 
67034  Strasbourg Cedex 2, France}
\author{Dietmar\ Weinmann}
\affiliation{Institut de Physique et Chimie des Mat\'eriaux de Strasbourg, 
UMR 7504 (CNRS-ULP), 23 rue du Loess, BP 43, 
67034  Strasbourg Cedex 2, France}
\author{Rodolfo\ A.\ Jalabert}
\affiliation{Institut de Physique et Chimie des Mat\'eriaux de Strasbourg, 
UMR 7504 (CNRS-ULP), 23 rue du Loess, BP 43, 
67034  Strasbourg Cedex 2, France}

\begin{abstract}
We propose a mechanism to describe spin relaxation in n-doped III-V 
semiconductors close to the Mott metal-insulator transition. Taking into 
account the spin-orbit interaction induced spin admixture in the hydrogenic 
donor states, we build a tight-binding model for the spin-dependent impurity 
band. Since the hopping amplitudes with spin flip are considerably smaller than 
the spin conserving counterparts, the resulting spin lifetime is very large. We 
estimate the spin lifetime from the diffusive accumulation of spin rotations 
associated with the electron hopping. Our result is larger but of the same 
order of magnitude than the experimental value. Therefore the proposed 
mechanism has to be included when describing spin relaxation in the impurity 
band. 
\end{abstract}

\maketitle

%\section{Introduction}
%\label{sec:intro}

The renewed interest in spin relaxation in semiconductors \cite{mei-zak} stems 
from the possible applications and fundamental science associated with the 
emerging field of spintronics \cite{aws-los-sam,zut-fab-das}. The large 
measured values of the electron spin lifetime constitute a promise for the use
of spin as a unit of quantum information and pose a considerable challenge for 
the identification of the appropriate mechanisms of spin relaxation. 
Interestingly, low-temperature experiments in various n-doped semiconductor 
bulk systems have found that the longest spin lifetimes occur close to the Mott
metal-insulator transition (MIT) density \cite{ale-hol,zar-cas,kik-aws,dzh}.
Many aspects of the Mott transition have been thoroughly studied, making it
a paradigm of Condensed Matter Physics \cite{mot}. However, the connection 
between the spin and transport problems is only beginning to be explored 
\cite{shk}.

The comprehensive experimental and theoretical work of Dzhioev \textit{et al.} 
\cite{dzh} considered both sides of the MIT, which for GaAs occurs at 
$n_\mathrm{c}\simeq 2\times 10^{16}\,\mbox{cm}^{-3}$. In the deeply localized 
regime, with donor densities $n \leq 5 \times 10^{15}\,\mbox{cm}^{-3}$, the 
hyperfine interaction was shown to account for the measured spin lifetimes 
\cite{dya-per-hyp}. For higher densities, but still lower than $n_\mathrm{c}$, 
the anisotropic exchange of localized spins was proposed as the dominant 
mechanism for spin relaxation \cite{kav}. Later calculations based on the same 
mechanism \cite{gor-kro} initiated an ongoing controversy concerning the 
quantitative agreement between experiment and theory.

For densities above the transition, the well-known D'yakonov-Perel (DP)
mechanism for conduction electrons was invoked \cite{dya-per}. This mechanism 
arises from the splitting of the conduction band due to spin-orbit interaction 
and yields a spin lifetime inversely proportional to the momentum relaxation 
time. This description applies to doping densities high enough so that mainly 
the conduction band is populated. (For GaAs the hybridization of impurity and 
conduction bands occurs at a doping density 
$n_\mathrm{h}\simeq 8\times 10^{16}\,\mbox{cm}^{-3}$ 
[\onlinecite{mat-toy,ale-hol,ser-gha}].)
If one is interested in understanding the large lifetimes measured near the 
MIT, the DP mechanism is not applicable, since, clearly, any mechanism invoking
momentum relaxation via impurity scattering becomes meaningless in the impurity 
band. This difficulty lies at the origin of the lack of suitable theories of 
spin relaxation at low temperature near the MIT \cite{footnote1}.

In this work we provide a theoretical frame and propose a spin-relaxation
mechanism for the metallic-conduction regime of the impurity band. Our approach 
is to extend, by incorporating the spin-orbit interaction, the well-known model 
of Matsubara and Toyozawa (MT) of electron conduction at zero temperature 
\cite{mat-toy}. The effect of the spin-orbit interaction is to introduce spin 
admixture in the impurity states. We will refer to this as impurity spin 
admixture (ISA). The ISA allows for spin-flip processes in electron-hopping 
events even in the absence of spin-dependent potentials. A tight-binding model 
built on the ISA states provides a theoretical framework to study spin dynamics 
and spin-dependent transport in the impurity band. Since we are interested in 
spin lifetimes, a first test of our model is to estimate the order of magnitude 
that the ISA mechanism predicts. We proceed by calculating the accumulated spin 
rotation angle along the diffusive evolution of the electron in the potential 
of the impurities. The time that it takes the spin to depart an angle of unity 
from its initial orientation is then taken as a qualitative measure of the spin
lifetime. 

The MT model consists of a tight-binding approximation built from the ground 
state of the doping impurities. For shallow donors it is a standard 
approximation to restrict the expansion of the impurity ground state to 
conduction-band state \cite{yu-car} and for an impurity located at the origin 
we write it as
\begin{equation} \label{eq:imp_trad}    
\left[\psi_{0,\sigma}\right]\!(\mathbf{r}) = 
\sum_{\mathbf{k}}      \phi(\mathbf{k}) \, 
e^{i \mathbf{k} \cdot \mathbf{r}} \,      
\left[u_{\mathbf{k},\sigma}\right]\!(\mathbf{r}) 
\approx      \phi(\mathbf{r}) \, 
\left[u_{\sigma}^{(0)}\right]\!(\mathbf{r}). 
\end{equation}
The envelope function $\phi(\mathbf{r})=(1/\pi a^3)^{1/2} \exp{(-r/a)}$, where 
$a$ is the effective Bohr radius ($a \approx 100 \mbox{\AA}$ for GaAs), is 
hydrogen-like. We note as $\phi(\mathbf{k})$ its Fourier transform, while 
$\left[u_{\mathbf{k},\sigma}\right](\mathbf{r})$ represents the periodic part of 
the Bloch functions of the conduction band states. Its dependence on 
$\mathbf{k}$, being much smoother than that of $\phi(\mathbf{k})$, leads to the 
last relation in Eq.\ (\ref{eq:imp_trad}), where we note 
$\left[u_{\sigma}^{(0)}\right]=\left[u_{\mathbf{k}=0,\sigma}\right]$. The spinors 
$\left[\psi_{0,\sigma}\right]$ and $\left[u_{\mathbf{k},\sigma}\right]$ are 
trivial since they are eigenstates of the operator $S_z$ with eigenvalue 
$\sigma = \pm 1$. However, this will no longer be the case once we include the 
spin-orbit interaction. The Hamiltonian of the MT model can be simply written as
\begin{equation}\label{eq:MT}
H_0 = \sum_{m\neq m',\sigma} t_{m m'}^{\sigma \sigma} \
c_{m' \sigma}^{\dag} \ c_{m \sigma},
\end{equation}
where $c_{m' \sigma}^{\dag}$ ($c_{m \sigma}$) represents the creation
(annihilation) of an impurity eigenstate at the impurity site $m'$ ($m$). The 
ground-state energy of an isolated impurity is taken as the origin of energies. 
The energy integral for the electronic transfer from site $m$ to $m'$ is given 
by
\begin{eqnarray}\label{eq:t_MT}
t_{m m'}^{\sigma \sigma} &\approx& 
\langle \psi_{m'\sigma}|V_{m'}|\psi_{m\sigma}\rangle \nonumber \\ 
&=& -V_0 \left(1+\frac{r_{mm'}}{a} \right) \exp{\left(-\frac{r_{mm'}}{a}\right)}.
\end{eqnarray}
The Coulomb-like potential produced by the impurity placed at $\mathbf{r}_m$
is $V_m(\mathbf{r})=-e^2 /\varepsilon |\mathbf{r}-\mathbf{r}_m|$. We note
$\varepsilon$ the static dielectric constant (12.9 for GaAs),
$V_0=e^2 /\varepsilon a$, $e$ the electron charge, and $r_{mm'}$ the distance 
between the two impurities. For convenience, in Eq.\ (\ref{eq:t_MT}) we 
switched from spinor to ket notation. The Hamiltonian Eq.\ (\ref{eq:MT}) has 
been thoroughly studied using a variety of analytical and numerical techniques 
\cite{mat-toy,chi-hub,pur-oda,gib-log-mad}, allowing a useful description of 
the impurity band and its electronic transport.

%\section{Tight-binding model with impurity spin admixture}
%\label{sec:ISA_model}

In order to extend the MT model to the spin case we first generalize the
shallow-donor wave functions to include the spin-orbit interaction. The 
spin-orbit effects coming from the orbital motion do not modify in an
appreciable way the envelope functions $\phi(r)$. Therefore, the spin-orbit 
interaction affects mainly the spinor $[u_{\mathbf{k}}]$. As is well-known, in 
lattices lacking inversion symmetry (i.e.\ zincblende semiconductors like GaAs),
the spin-orbit coupling leads to spin-mixed conduction-band states. 
Group-theoretical arguments dictate the way in which the conduction and valence 
states are mixed by the spin-orbit interaction. Within the 
$\mathbf{k} \cdot \mathbf{p}$ approximation of Kane \cite{kan} (and using the 
notation of Ref.\ \onlinecite{cha}), the periodic part of the spin-mixed 
conduction-band states is given by
\begin{equation}\label{eq:spin-mixed}
|\tilde{u}_{\mathbf{k}\sigma} \rangle = 
|u_{\sigma}^{(0)} \rangle +   \mathbf{k} \cdot |\mathbf{u}_{\sigma}^{(1)} \rangle,
\end{equation}
where
\begin{equation}
|\mathbf{u}_{\sigma}^{(1)} \rangle = \alpha_1 \left(|\mathbf{R}\sigma\rangle 
+ \alpha_2 \mathbf{S} \times |\mathbf{R}\sigma\rangle \right).
\end{equation}
The state $|u_{\sigma}^{(0)}\rangle$ is $s$-like, since it describes the
unperturbed wave function at the $\Gamma$-point. The vector
$|{\mathbf R}\rangle = (|X\rangle, |Y\rangle, |Z\rangle)$ represents the three 
$p$-like valence states. ${\mathbf S}$ is the spin operator. The state 
$|\tilde{u}_{\mathbf{k}\sigma} \rangle$ is clearly not an eigenstate of $S_z$. 
However, we still characterize it with the label $\sigma$ since the mixing is 
small, and 
$\langle\tilde{u}_{\mathbf{k}\sigma}|S_z|\tilde{u}_{\mathbf{k}\sigma}\rangle$
is much closer to $\sigma \hbar/2$ than to $-\sigma \hbar/2$. The weak spin 
mixing is governed by the small constants 
$\alpha_1 = i\hbar \left[(3E_G+2\Delta)/(6m^*E_G(E_G+\Delta))\right]^{1/2}$ and 
$\alpha_2 = 2\Delta / i\hbar(2\Delta+3E_G)$. We note $\Delta$ the spin-orbit 
splitting of the valence bands, $m^*$ the conduction-band effective mass, 
and $E_G$ the bandgap.
 
The mixing of Eq.\ (\ref{eq:spin-mixed}) tells us that in the presence
of SO interaction, Eq.\ (\ref{eq:imp_trad}) takes the form
\begin{equation}  
\left[\tilde{\psi}_{0\sigma}\right]\!\!(\mathbf{r}) =   
\sum_{\mathbf{k}} \phi(\mathbf{k}) \, e^{i \mathbf{k} \cdot \mathbf{r}} \,  
\left(\left[u_{\sigma}^{(0)}\right]\!\!(\mathbf{r}) +         
\mathbf{k} \cdot \left[\mathbf{u}_{\sigma}^{(1)}\right]\!\!(\mathbf{r}) \right).
\end{equation}
Using the hydrogenic character of $\phi(r)$, the impurity spin-admixture
(ISA) state centered at $\mathbf{r}_m$ reads 
\begin{eqnarray}\label{eq:imp_spin}
\left[\tilde{\psi}_{m\sigma}\right]\!\!(\mathbf{r}) &=& 
\phi(\mathbf{r}-\mathbf{r}_m)  \\ \nonumber  
&\times & \left(\left[u_{\sigma}^{(0)}\right]\!\!(\mathbf{r}) +    
\frac{i}{a} \,  \frac{(\mathbf{r}-\mathbf{r}_m)}{|\mathbf{r}-\mathbf{r}_m|} 
\cdot \left[\mathbf{u}_{\sigma}^{(1)}\right]\!\!(\mathbf{r}) \right).
\end{eqnarray}

Electron hopping between ISA states in different impurity sites provides a 
mechanism for spin-flip by connecting the $\sigma$ and 
$\overline{\sigma}=-\sigma$ states. The Hamiltonian of the system now contains 
a term without spin-flip (described by the Hamiltonian $H_0$ of 
Eq.\ (\ref{eq:MT})) and a spin-flip term
\begin{equation}\label{eq:ISA}
H_1 = \sum_{m\neq m',\sigma} t_{m m'}^{\sigma \overline{\sigma}} \
c_{m' \overline{\sigma}}^{\dag} \ c_{m \sigma}.
\end{equation}
As in the spinless case, the matrix element 
$\langle\tilde{\psi}_{m' \sigma}|V_{m'}|\tilde{\psi}_{m \sigma}\rangle$ is 
expected to dominate the energy integral for the electronic transfer between 
states $m$ and $m'$. With spin flip, however, symmetry reasons dictate that the 
corresponding matrix element 
$\langle\tilde{\psi}_{m' \overline{\sigma}}|V_{m'}|\tilde{\psi}_{m \sigma}\rangle$
vanishes. This important fact is ultimately responsible for the large values 
of the spin lifetime in the regime of impurity-band conduction. We are then 
forced to consider the matrix element
\begin{eqnarray}\label{eq:3center}
&&\langle\tilde{\psi}_{m' \overline{\sigma}}|V_p|\tilde{\psi}_{m \sigma}\rangle =
-C \int d^3r \\ \nonumber 
&&\quad\quad\quad \times \frac{(r-r_m)_-(z-z_{m'})-(z-z_m)(r-r_{m'})_-}
{|\mathbf{r}-\mathbf{r}_m|\,|\mathbf{r}-\mathbf{r}_p|\,
|\mathbf{r}-\mathbf{r}_{m'}|} \\ \nonumber 
&&\quad\quad\quad \times \exp{\left(-\frac{|\mathbf{r}-\mathbf{r}_m|
+|\mathbf{r}-\mathbf{r}_{m'}|}{a}\right)}, 
\end{eqnarray}
with $p\neq m,m'$, $r_{\pm}=x\pm i y$, $C=V_0 |\alpha_1|^2\alpha_3 /\pi a^4$, 
and $\alpha_3 = 3\Delta (\Delta + 2E_G)/(2\Delta + 3E_G)^2$. Three-center 
integrals like the one of Eq.\ (\ref{eq:3center}) are in general very difficult 
to calculate \cite{sla}. However, in the case $r_{mm'}/a \gg 1$ 
[\onlinecite{footnote2}] we can perform the integrals using the saddle-point 
approximation and obtain
\begin{eqnarray}\label{eq:mat_elem}
\langle \tilde{\psi}_{m'\overline{\sigma}}|V_p|\tilde{\psi}_{m\sigma} \rangle 
&=& -4.2\, C\, e^{i\varphi_m} \, (\cos{\phi_p} + i\, \cos{\theta_m}\sin{\phi_p}) 
\nonumber \\
&\times&  \frac{r_{mm'}^{3/2} a^{1/2}\rho_p \, \exp{(-r_{mm'}/a)}} 
{\left(1+(\rho_p^2+z_p^2) r_{mm'}/a \right)^{3/2}}.
\end{eqnarray}
$\varphi_m$ and $\theta_m$ are the polar angles of the vector $\mathbf{r}_{mm'}$
in the original coordinate system. $\phi_p$, $\rho_p$, and $z_p$ are the 
cylindrical coordinates of $\mathbf{r}_p$ in a new system having the z-axis 
parallel to $\mathbf{r}_{mm'}$ and its origin at the middle point between the 
sites $m$ and $m'$. For a given impurity configuration, the matrix elements 
(\ref{eq:mat_elem}) yield the energy integral with spin flip 
\begin{equation}\label{eq:t_spin_flip}
t_{m m'}^{\sigma \overline{\sigma}} = \sum_{p\neq m,m'}
\langle \tilde{\psi}_{m' \overline{\sigma}}|V_p|\tilde{\psi}_{m \sigma}\rangle
\end{equation}
of the Hamiltonian $H_1$ (Eq.\ (\ref{eq:ISA})). The system Hamiltonian $H_0+H_1$
can be addressed numerically or by perturbation theory. In order to test the 
physical relevance of the proposed spin-flip mechanism, we will estimate the 
spin relaxation time within some simplifying hypothesis that we discuss in what 
follows.

%\section{Spin relaxation time: diffusive approach}
%\label{sec:dif_app}

Viewing the electron transport as a hop between impurity sites, we see that,
since $|t_{m m'}^{\sigma \sigma}| \gg |t_{m m'}^{\sigma \overline{\sigma}}|$, there 
is a very small probability of spin-flip per hop, which may be translated into 
a mean spin-rotation angle $\gamma_{m m'}$. Assuming that these relative 
rotations are accumulated in a diffusive way, we can estimate the 
spin-relaxation time by
\begin{equation}\label{eq:taus}
\frac{1}{\tau_\mathrm{s}} = \frac{2}{3}\frac{<\gamma^2>}{\tau_\mathrm{c}},
\end{equation}
where $<\gamma^2>$ is the ensemble average of $\gamma_{m m'}^2$ and 
$\tau_\mathrm{c}$ is the mean hopping time. If in the hop between impurities $m$
and $m'$ the electron is initially in the spin-up state, the expectation value 
of the spin operator after the hop is
\begin{equation}
<\mathbf{S}_{m'}> = 
\frac{1}{|t_{m m'}^{\sigma \sigma}|^2+|t_{m m'}^{\sigma \overline{\sigma}}|^2}  
\left( \begin{array}{c}  
2\,\mbox{Re}[t_{m m'}^{\sigma \sigma} \, t_{m m'}^{\sigma \overline{\sigma}}] \\  
2\,\mbox{Im}[t_{m m'}^{\sigma \sigma} \, t_{m m'}^{\sigma \overline{\sigma}}]\\     
|t_{m m'}^{\sigma \sigma}|^2 - |t_{m m'}^{\sigma \overline{\sigma}}|^2           
\end{array} \right).
\end{equation}
The angle between the initial and final spin orientations is
\begin{equation}\label{eq:gamma}
\theta_{m m'}=  \mbox{Arccos} 
\left(\frac{|t_{m m'}^{\sigma \sigma}|^2-|t_{m m'}^{\sigma \overline{\sigma}}|^2}
{|t_{m m'}^{\sigma \sigma}|^2+|t_{m m'}^{\sigma \overline{\sigma}}|^2}\right) \simeq 
\frac{2 |t_{m m'}^{\sigma \overline{\sigma}}|}{|t_{m m'}^{\sigma \sigma}|}.
\end{equation}
Allowing for an arbitrary initial spin orientation before the hop enhances the 
root mean square value of $\theta_{m m'}$ by a factor of $\sqrt{3/2}$. We then 
have $<\gamma_{mm'}^2> = (3/2)^2 {<\theta_{mm'}^2>}$, where an additional factor 
of $3/2$ appears since $\theta_{m m'}$ contains the two components of 
$\gamma_{mm'}$ that are relevant for spin relaxation. 

Given the form of (\ref{eq:t_spin_flip}) of the energy integrals, the typical
rotation angle $\gamma_{m m'}^2$ involves a double sum over impurities $p$ and 
$p'$ ($\neq m, m'$), which can be approximated by its impurity average. Only 
the diagonal term ($p=p'$) survives the average, yielding
\begin{equation}\label{eq:gcmmp} 
\gamma_{m m'}^2 = 33.2\,\left(\frac{C}{V_0}\right)^2 \, 
r_{mm'}^{3/2} \, a^{11/2} \, n_\mathrm{i} \, (1+\cos^2\theta_{mm'}),
\end{equation} 
where $n_\mathrm{i}$ is the impurity density. The typical rotation angle between
impurities $m$ and $m'$ increases with their distance $r_{mm'}$ as a power law. 
This dependence renders the impurity distribution crucial for the determination 
of the mean square rotation angle per hop. The distribution of doping impurities
is known to be completely random and to lack hard-core repulsive effects on the 
scale of $a$ [\onlinecite{tho-etal}]. Since the probability of jumping from a 
given impurity $m$ to a second one $m'$ is 
$|t_{m m'}^{\sigma \sigma}|^2/\sum_{m'} |t_{m  m'}^{\sigma \sigma}|^2$,
we obtain the impurity average of the typical rotation angle per hop as
\begin{equation} \label{eq:gcprom} 
<\gamma^2> = 1.8\times 10^2\, \left(\frac{C}{V_0}\right)^2 a^4
\left({n_\mathrm{i} a^3}\right).
\end{equation}

The hopping time can be estimated from perturbation theory by determining the 
characteristic time for the decay of the initial population by one-half, 
yielding
\begin{equation}\label{eq:tauc}
\frac{1}{\tau_\mathrm{c}}=\frac{1}{\hbar}
\sqrt{2 \sum_{m'}{|t_{m m'}^{\sigma \sigma}|^2}},
\end{equation}
where an average over the initial position $m$ is implicit. This estimation 
assumes orthogonality of the electronic orbitals at different impurity sites, 
which is actually not satisfied by the hydrogenic states. However, since the 
overlaps are very small, the non-orthogonality effects arising in the MT model 
are known to be small \cite{maj-and,chi-hub}.

We calculate the rotation angle for hopping processes between impurity sites. 
However, if we take delocalized impurity-band initial states, a golden rule 
approach with some simplifying assumptions leads to the same expression for 
$\tau_\mathrm{s}$ up to a factor of order one \cite{twj_unpublished}.

From Eqs.\ (\ref{eq:taus}), (\ref{eq:gcprom}), and (\ref{eq:tauc}) we obtain
\begin{equation}\label{eq:taus_final}
\frac{1}{\tau_\mathrm{s}} = 8.2\times 10^2 \, \frac{C^2 a^4}{V_0 \hbar} 
\left({n_\mathrm{i} a^3}\right)^{3/2}.
\end{equation}
For GaAs, at the density of the MIT, Eq.\ (\ref{eq:taus_final}) yields
a spin-relaxation time of $\tau_\mathrm{s} = 1200$~ns.

This value is larger but within an order of magnitude of the experimentally 
reported result of 200~ns [\onlinecite{kik-aws,dzh}] at the MIT. We stress that 
our result does not depend on any adjustable parameter, but it relies on a few 
approximations. For example, a step of the calculation that could admit an 
alternative treatment is the impurity average of $\gamma_{mm'}^2$. If we assume 
that the hopping takes place only between nearest neighbors (separated by a 
typical distance $n_\mathrm{i}^{-1/3}$) we obtain a value of 700~ns for 
$\tau_\mathrm{s}$ at the MIT. In general, the order of magnitude agreement with 
the experimental value is not affected by the approximations and therefore the 
ISA mechanism needs to be included in the description of spin relaxation in the 
impurity band.

In materials with stronger spin-orbit interaction like InSb and InAs, one 
obtains considerably smaller values of the spin lifetime. For the impurity
density of $5 \times 10^{14} \mbox{cm}^{-3}$ in InSb, our estimate yields a spin
lifetime of 86~ns. This value is within an order of magnitude of the 
experimental result and close to the theoretical estimation of Ref.\ 
\onlinecite{cha}. The theoretical approach of this reference is justified only 
in the high-concentration limit since it describes single scattering of plane 
waves and treats electron-electron interactions through a corrective factor.

%\section{Conclusion}
%\label{sec:conclusion}

In this work we have proposed a new mechanism for spin relaxation in the regime 
where electron conduction occurs in the impurity band of doped semiconductors. 
The mechanism is based on the impurity spin admixture of the electronic ground 
state of the donors caused by the spin-orbit interaction. The impurity spin 
admixture states do not have a well-defined spin projection about a fixed 
spatial direction, and therefore hopping between two of these states may 
connect different projections of the angular momentum. 

Unlike the spinless case, the matrix elements of the spin-flip hops are not
dominated by a two-center integral, where the impurity potential corresponds to 
one of the extreme sites. We therefore have to consider three-center integrals, 
where the impurity potential is not centered around any of the two sites of the 
hop. Since the latter matrix elements are considerably reduced with respect to 
the former, the resulting spin lifetimes are very large.

Our calculation of spin-flip matrix elements yields a suitable model for 
studying electron and spin transfer in the regime of impurity concentrations
just above that of the metal-insulator transition. Various treatments can be
applied to our model Hamiltonian. In this work we provide an estimation of the 
spin lifetime by calculating the diffusive accumulation of spin rotation during 
the hopping process. This estimation yields values that are larger than the 
ones experimentally measured but within the right order of magnitude. Therefore 
the impurity spin admixture mechanism has to be taken into account in 
descriptions of the spin relaxation in the impurity band. Our model admits 
generalizations including other physical effects, like doping compensation, 
electron-electron interaction, and a second electronic band, which may improve 
the agreement between theory and experiment.

\begin{acknowledgments}
We are grateful to M. Sanquer for useful discussions and directing us to
Ref.~\onlinecite{tho-etal}. We acknowledge financial support from CONICET 
(PIP-5851), UBACyT (X179), and from the European Union through the MCRTN 
program. P.I.T. is a researcher of CONICET. 
\end{acknowledgments}


\begin{thebibliography}{99}

\bibitem{mei-zak}
{\it Optical Orientation}, edited by F.\ Meier and B.\ Zakharchenya,
Vol.\ 8 of Modern Problems in Condensed Matter Sciences (North-Holland, Amsterdam, 1984).

\bibitem{aws-los-sam} 
{\it Semiconductor Spintronics and Quantum Computation}                      
D.\ D.\ Awschalom, D.\ Loss, N.\ Samarth (eds) 
(Springer, Berlin, 2002).

\bibitem{zut-fab-das} I.\ Zuti\'c, J.\ Fabian, and S.\ Das Sarma, 
Rev.\ Mod.\ Phys.\ {\bf 76}, 323 (2004).

\bibitem{ale-hol}
M.\ N.\ Alexander and D.\ F.\ Holcomb, 
Rev.\ Mod.\ Phys.\ {\bf 40}, 815 (1968).

\bibitem{zar-cas}
V.\ Zarifis and T.\ G.\ Castner, 
Phys.\ Rev.\ B {\bf 36}, 6198 (1987).

\bibitem{kik-aws}
J.\ M.\ Kikkawa and D.\ D.\ Awschalom,
Phys.\ Rev.\ Lett.\ {\bf 80}, 4313 (1998).

\bibitem{dzh}
R.\ I.\ Dzhioev, K.\ V.\ Kavokin, V.\ L.\ Korenev, M.\ V.\ Lazarev, 
B.\ Y.\ Meltser, M.\ N.\ Stepanova, B.\ P.\ Zakharchenya, D.\ Gammon, 
and D.\ S.\ Katzer, 
Phys.\ Rev.\ B {\bf 66}, 245204 (2002).

\bibitem{mot}
{\it Conduction in non-crystalline materials}, by Sir Nevill Mott 
(Oxford Science Publications, 
Clarendon Press, Oxford, 1987).

\bibitem{shk} 
B.\ I.\ Shklovskii, 
Phys.\ Rev.\ B {\bf 73}, 193201 (2006).

\bibitem{dya-per-hyp} 
M.\ I.\ D'yakonov and V.\ I.\ Perel, 
Sov.\ Rev.\ JETP {\bf 38}, 177 (1974).

\bibitem{kav} 
K.\ V.\ Kavokin, 
Phys.\ Rev.\ B {\bf 64}, 075305 (2001).

\bibitem{gor-kro} 
L.\ P.\ Gor'kov and P.\ L.\ Krotkov, 
Phys.\ Rev.\ B {\bf 68}, 155206 (2003).

\bibitem{dya-per}
M.\ I.\ D'yakonov and V.\ I.\ Perel, 
Sov.\ Rev.\ JETP {\bf 33}, 1053 (1971); 
Sov.\ Phys.\ Solid State {\bf 13}, 3023 (1972).

\bibitem{mat-toy}
T.\ Matsubara and Y.\ Toyozawa, 
Prog.\ Theoret.\ Phys.\ {\bf 26},739 (1961).

\bibitem{ser-gha}
J.\ Serre and A.\ Ghazali,
Phys.\ Rev.\ B {\bf 28}, 4704 (1983).

\bibitem{footnote1} This limitation is clearly presented in the theoretical 
survey of Ref.\ \onlinecite{dzh}. For temperatures above 50~K, a fourteen-band 
calculation of momentum relaxation using the concept of motional narrowing 
approaches the experimental values for bulk GaAs \cite{lau-ole-fla}. 

\bibitem{yu-car}
{\it Fundamentals of Semiconductors}, 
by P.\ Y.\ Yu and M.\ Cardona (Springer-Verlag, Berlin, 2001).

\bibitem{chi-hub} 
W.\ Y.\ Ching and D.\ L.\ Huber, 
Phys.\ Rev.\ B{\bf 26}, 5596 (1982).

\bibitem{pur-oda} 
A.\ Puri and T.\ Odagaki, 
Phys.\ Rev. B, {\bf 29},1707 (1984).

\bibitem{gib-log-mad} 
M.\ K.\ Gibbons, D.\ E.\ Logan, and P.\ A.\ Madden, 
Phys.\ Rev. B, {\bf 38}, 7292 (1988).

\bibitem{kan}
E.\ O.\ Kane, 
J.\ Phys.\ Chem.\ Solids {\bf 1}, 249 (1957).

\bibitem{cha} 
J.-N.\ Chazalviel, 
Phys.\ Rev.\ B {\bf 11}, 1555 (1975).

\bibitem{sla} 
{\it Quantum Theory of Molecules and Solids}, 
Vol.\ 1,J.\ C.\ Slater (McGraw-Hill Book Company, 1963).

\bibitem{footnote2} At the MIT we have $r_{mm'}/a=3.7$, while at the maximum 
density we consider $r_{mm'}/a=2.3$.

\bibitem{tho-etal} 
G.\ A.\ Thomas, M.\ Capizzi, F.\ DeRosa, R.\ N.\ Bhatt, and T.\ M.\ Rice, 
Phys.\ Rev.\ B {\bf 23}, 5472 (1981).

\bibitem{maj-and} 
N.\ Majlis and E.\ Anda, 
J.\ Phys.\ C: Solid State Phys.\ {\bf 11}, 1607 (1978).

\bibitem{twj_unpublished}
P.\ I.\ Tamborenea, D.\ Weinmann, and R.\ A. Jalabert, unpublished.

\bibitem{lau-ole-fla}
W.\ H.\ Lau, J.\ T.\ Olesberg, and M.\ E.\ Flatt\'e, 
Phys.\ Rev.\ B {\bf 64}, 161301(R) (2001).

\end{thebibliography}
\end{document}